\documentclass{amsart}
\begin{document}
\title{ MULTIPLY WARPED PRODUCTS \\ WITH NON-SMOOTH METRICS}\par
\author {JAEDONG CHOI}
\address{\noindent Department of Mathematics P. O. Box 335-2 Air Force Academy,
 Ssangsu, Namil, Cheongwon, Chungbuk, South Korea 363-849}\vskip10pt
\email{\noindent {\it{E-mail address}}: jdong@afa.ac.kr  }
\begin{abstract} In this article we study manifolds with $C^{0}$-metrics and properties of Lorentzian multiply warped products. We represent the interior Schwarzschild space-time as a multiply warped product space-time with warping functions and we also investigate the curvature of a multiply warped product with $C^0$-warping functions.  We given the {\it{Ricci curvature}} in terms of $f_1$, $f_2$ 
for the multiply warped products of the form $M=(0,\ 2m)\times_{f_1}R^1\times_{f_2} S^2$.  
\end{abstract}
\maketitle
\vskip18pt
\noindent{\bf I. INTRODUCTION}
\vskip10pt
The concept of a warped product manifold was introduced by Bishop and O'Neill${}^1$, where it served to provide a class of complete Riemannian manifolds with everywhere negative curvature. The connection with general relativity was first made by Beem, Ehrich, and Powell${}^{2,3}$, who pointed out that several of the well-known exact solutions to Einstein's field equations are pseudo-Riemannian warped products. Beem and Ehrich${}^{4}$ further explored the extent to which certain causal and completeness properties of a space-time maybe determined by the presence of a warped product structure. O'Neill's book on Semi-Riemannian geometry${}^{5}$ took this line of development to a natural conclusion by elevating warped products to a central role. After first developing the general theory of warped products to spaces, O'Neill then applied the theory to discuss, in turn, the special cases of Robertson-Walker, Friedmann, static and Schwarzshild space-time. The role of warped products in the study of exact solutions to Einstein's equations is now firmly established, and it appears that these structures are generating interest in other areas of geometry.\vskip8pt
We study manifolds with $C^{0}$-metrics 
and properties of Lorentzian multiply warped products. Many authors${}^{6, 7, 8, 9}$ have dealt with Lorentzian manifolds with non-smooth metric tensors from various view points. Of particular interest are space-times which have a metric tensor which is fails to be $C^{1}$ across a hypersurface, and is $C^{\infty}$ off the hypersurface. A space-time which, in an admissible coordinate system, the metric tensor is continuous but has a jump in its first and second derivatives across a submanifold will have a curvature tensor containing a Dirac delta function${}^{10,11}$. The support of this distribution may be of three, two, or one dimensional or may even consist of a single event. Lichnerowicz's formalism${}^{12}$ for dealing with such tensors is modified so as to obtain a formalism in which the Riemannian curvature tensor and Ricci curvature tensor exist in the sense of distributions. We note that Smoller and Temple${}^{13}$ have presented a general theory for matching two solutions of the Einstein field equations at arbitrary shock-wave interface across which the metric $g$ is $C^0$-Lorentzian, i.e., smooth surface across which the first derivatives of the metric suffer at worst a jump discontinuity. 
A {\it multiply warped products space-time} with base $(B,-dt^2)$, fibers $(F_i,g_i)$ $i=1,...,n$ and warping functions $f_i>0$ is the product manifold $(B\times F_1\times...\times F_n, g)$ endowed with the Lorentzian metric:
$g=-\pi_B^{\ast}dt^2+\sum_{i=1}^n(f_i\circ\pi_B)^2\pi_i^\ast g_i\equiv-dt^2+\sum_{i=1}^nf_i^2g_i$
where $\pi_B$, $\pi_i$\ $i=1,...,n$ are the natural projections of $B\times F_1\times...\times F_n$ onto $B$ and $F_1$,...,$F_n$, respectively. Thus, warped product 
spaces are extended to a richer class of spaces involving multiply products. Multiply warped product spaces were studied by Flores, J, L. and M. S\'{a}nchez${}^{14}$.

 The conditions of spacelike boundaries in the multiply warped product spacetimes were studied by Steven G. Harris${}^{15}$. From a physical point of view, these space-time are interesting, first, because they include classical examples of space-time: when $n=1$ they are {\it Generalized Robertson-Walker} space-times, standard models of cosmology; when $n=2$, the intermediate zone of Reissner-Norsdstr\"{o}m space-time and interior of Schwarzschild space-time appear as particular cases${}^{16}$. \vskip5pt
The conditions of spacelike boundaries in the multiply warped product spacetimes were studied by Steven G. Harris. The Kasner metric was studied as a cosmological model by Sch\"{u}cking and Heckmann${}^{17}$(1958)
\vskip5pt The Schwarzschild solution is interpreted as describing the gravitational field of a point particle with mass $m$. Generally, this metric is taken to represent the outside metric for a star with $r>r_0$ where $r_0$ gives the boundary of the star. The metric inside $r<r_0$ is a different interior metric determined by the matter distribution $T_{ij}$ inside the star and is matched at the boundary $r=r_0$ with the metric. We represent the interior Schwarzschild space-time as a multiply warped product space-time with warping functions and we also investigate the curvature of a multiply warped product with $C^0$-warping functions. We given the {\it{Ricci curvature}} in terms of $f_1$, $f_2$ 
for the multiply warped products of the form $M=(0,\ 2m)\times_{f_1}R^1\times_{f_2} S^2$. 

\vskip20pt \noindent{\bf II. MULTIPLY WARPED PRODUCTS MANIFOLDS\\ WITH NON-SMOOTH METRIC TENSOR}
\vskip10pt
 
In this section, we state some definitions and standard results${}^{2,5}$ which will be needed below. A Lorentzian manifold $(M,g)$ is a connected smooth manifold of dimension  ${\geq2}$ with a countable basis together with a Lorentzian metric \ $g$ \ of signature $(-, +, +,...,+)$.
Let $(F_i, g_i)$ be Riemannian manifolds, and let $(B, g_B)$ be either a spacetime, or let $B$ be $R^1$ with $g_B=-dt^2$. Let $f_i>0$, $i=1,...,n$ be smooth functions on $B$. A {\it multiply warped products space-time} with base $(B, g_B)$, fibers $(F_i,g_i)$ $i=1,...,n$ and warping functions $f_i>0$ is the product manifold $(B\times F_1\times...\times F_n, g)$ endowed with the Lorentzian  metric:
$$g=\pi_B^{\ast}g_B+\sum_{i=1}^n(f_i\circ\pi_B)^2\pi_i^\ast g_i\equiv-dt^2+\sum_{i=1}^nf_i^2g_i\eqno(2.1)$$
where $\pi_B$, $\pi_i$\ $i=1,...,n$ are the natural projections of $B\times F_1\times...\times F_n$ onto $B$ and $F_1$,...,$F_n$, respectively.\vskip10pt
{\bf Definition 2.1}\ \  Suppose $M=B\times_{f_1} F_1\times...\times_{f_n} F_n$ is a multiply warped product of semi-Riemannian manifolds. For any $(p,q_1,...,q_n)\in M$, the submanifold $B\times q_1\times...\times q_n$ is a leaf of $M$. The submanifolds $p\times F_1\times...\times q_n$ is called type 1 fibers and $p\times q_1\times...\times F_n$ is called type n fibers, respectively.
As in the case of warped products, a multiply warped metric satisfies:\par
(1) For each $q_i\in F_i$, $\left.{\pi}_B\right|_{(B\times q_1\times...\times q_n)}$ is an isometry onto $B$\par
(2) For each $p\in B$ and $q_j\in F_j$,  $\left.{\sigma_i}\right|_{(p\times q_1\times...\times F_i\times...\times q_n)}$ is homothety onto $F_i$, with scale factor $1/f_i(p)$.\par

(3) For each $(p, q_1,..., q_n)\in M$, the set $B\times q_1\times...\times q_n$ and the $i$-th type fiber, $p\times q_1\times...\times F_i\times...\times q_n$ are mutually orthogonal at $(p, q_1,...,q_n)$.\vskip8pt

 Let $M$ be a smooth manifold of dimension $n$.
Then subset $S$ of $M$ is a regularly embedded hypersurface of $M$ if for all $p\in S$, there exists a coordinate neighborhood $U(p)$ with coordinates $(x_{1},...,x_{n})$ such that $S\cap U = \{(x_{1},...,x_{n})\in U\mid x_{n}=p\}$. For convenience, we say that such a  
neighborhood $U$ is partitioned by $S$. We denote $\{x \in U\mid x_{n}>p\}$
and  $\{x \in U\mid x_{n}<p\}$ by $U_{p}^{+}$ and $U_{p}^{-},$ respectively.
Now let $M$ be a smooth manifold with a regularly embedded hypersurface $S$. Let $S^c$ denote the complement of $S$. We define the concept of a $C^{0}$-Lorentzian metric on $M$
\vskip15pt
{\bf Definition 2.2}\ \ A $C^{0}$-Lorentzian metric on $M$ is a nondegenerate (0,2) tensor of Lorentzian signature  such that:\par
\hskip0.4cm(1) $g \in C^{0}$ on $S$\par
\hskip0.4cm(2) $g \in C^{\infty}$ on $M\cap S^{c}$\par
\hskip0.4cm(3) For all $p \in S$, and $U(p)$ partitioned by $S$, $g{\mid}_{U_{p}^{+}}$ and $g{\mid}_{U_{p}^{-}}$ have smooth extensions to {\hskip2cm $U$}.\vskip8pt
We call $S$ a $C^{0}$-singular hypersurface of $(M, g)$. 
\vskip15pt
 \noindent{\bf III.  MULTIPLY WARPED PRODUCTS MANIFOLDS WITH $C^0$ WARPING FUNCTION}\vskip10pt
Let $M=B\times_{f_1} F_1\times...\times_{f_n} F_n$ be a multiply warped products with $g_B=-dt^2$. Let $f_i>0$ $i=1,...,n$ smooth functions on $B=(a, b)$. Recall $f_i\in C^{\infty}$ for $t\not=t_0$ if $f_i\in C^0$ at $t=p$. When $f\in C^0$ at a single point $p\in R^1$ and $S=\{p\}\times_{f_1} F_1\times...\times_{f_n} F_n$, we use the notation $f\in C^{0}(S)$.

\vskip15pt
{\bf Proposition 3.1}\ \ Let $M=B\times_{f_1} F_1\times...\times_{f_n} F_n$ be a multiply warped products with Riemannian curvature tensor $R$. If $X$, $Y\in \frak{L}(B)$, $V_i$, $W_i\in\frak{L}(F_i)$ where $1\leq i\leq n$, $f_i\in C^0(S)$,  then   \par
\hskip0.4cm  (1) $\nabla_{X}Y\in \frak L(B)$ is the lift of $\nabla_{X}Y$ on $\frak L(B)$,\par
\hskip0.4cm  (2) $\nabla_{X}V_i = \nabla_{V_i}X =(Xf_i/f_i)V_i=\biggl\{X^1{\left({f_i'}^{+}u(t-p)+{f_i'}^{-}u(p-t)\right)}/f_i\biggr\}V_i$.\par
\hskip0.4cm  (3)\ \  $\nabla_{V_i}{V_j}=\nabla_{V_j}{V_i}=0$ if $i\not=j$\par
\hskip0.4cm  (4) nor$\nabla_{V_i}W_i$= $II^i(V_i,W_i) $= $-$\ $(<V_i, W_i>/f_i)\thinspace {\text {grad}}\ {f_i}$ \par \hskip2.5cm  = $(<V_i, W_i>/f_i){\left({f_i'}^{+}u(t-p)+{f_i'}^{-}u(p-t)\right)}{\frac{\partial }{\partial t}}$. \par
\hskip0.4cm  (5) tan$\nabla_{V_i}W_i \in \frak L(F_i)$ is the lift of $\nabla_{V_i}W_i$ on $F$.
 \vskip10pt
{\it Proof}\par Refer to O'Neill${}^{5}$ results and similar arguments yield the (1), (3), (5).  

Clearly, it is only necessary to establish (2) and (4)\par
\hskip0.4cm (2) Since $X(f_i)=X^1{\left({f_i'}^{+}u(t-p)+{f_i'}^{-}u(p-t)\right)}/f_i$\ \ where $X=X^1{\frac{\partial}{\partial t}}$\par \hskip0.4cm  we have $(Xf_i/f_i)V_i=\biggl\{X^1{\left({f_i'}^{+}u(t-p)+{f_i'}^{-}u(p-t)\right)}/f_i\biggr\}V_i$.

\par
\hskip0.4cm (4) Since ${\text {grad}}\ {f_i}=\sum_{k=1}^{n}g^{kj}{\frac{\partial f_i}{\partial x^{k}}}{\frac{\partial }{\partial x^{j}}}
= -{\frac{\partial f_i}{\partial t}}{\frac{\partial }{\partial t}}$\par
\hskip3cm$=-\left({{f_i}'}^{+}u(t-p)+{{f_i}'}^{-}u(p-t)\right){\frac{\partial }{\partial t}}$\par
\hskip0.5cm${\text {nor}}\nabla_{V_i}W_i $= $II^i(V_i,W_i)$ 
\par\hskip2cm= {$-$} $(<V_i, W_i>/{f_i}){\text {grad}}\ {f_i}$
\par\hskip2cm=$(<V_i, W_i>/f_i){\left({{f_i}'}^{+}u(t-p)+{{f_i}'}^{-}u(p-t)\right)}{\frac{\partial }{\partial t}}$.
\qed
\vskip15pt
{\bf Proposition 3.2}\ \ Let $M=B\times_{f_1} F_1\times...\times_{f_n} F_n$ be a multiply warped products with Riemannian curvature tensor $R$. If $X$, $Y\in \frak{L}(B)$, $U_i$, $V_i$, $W_i\in\frak{L}(F_i)$ where $1\leq i\leq n$, $f_i\in C^0(S)$,  then   \par

\hskip0.4cm (1) $R_{XY}Z \in \frak L(B)$ is the lift of ${}^{B}R_{XY}Z(=0)$ on $\frak L(B)$,\par
 \hskip0.4cm  (2) $R_{XU_i}{U_j}=R_{U_iU_j}{X}=R_{{U_j}{X}}{U_i}=0$ for 
 $i\not=j$,\par
\hskip0.4cm  (3) $R_{U_iX}Y = (H^{f_i}(X, Y)/f_i)U_i$
\par\hskip2.2cm=$\biggl\{\biggl\{X^{1}Y^{1}\left(f_i''(t)+\delta_{p}(t)({f_i'}^{+}-{f_i'}^{-})\right)\biggr\}/f_i
\biggr\}U_i$,\par\hskip2.5cm
where $H^f_i$ is the Hessian of $f_i$.\par
 \hskip0.4cm  (3) $R_{XU_i}{U_j}=R_{U_iU_j}{X}=R_{{U_j}{X}}{U_i}=0$ for 
 $i\not=j$,\par
 \hskip0.4cm  (4) $R_{XY}U_i=0$ for each $i=1,..,n$,\par
 \hskip0.4cm (5) $R_{U_iV_i}U_j=0$\  for $i\not=j$,\par

 \hskip0.4cm (6) $R_{U_iU_j}V_j={\frac{<U_j,V_j>}{{f_if_j}}}\left({{f_j}'}^{+}+{{f_j}'}^{-}\right)\left({{f_i}'}^{+}+{{f_i}'}^{-}\right)U_i$\  for $i\not=j$,\par
\hskip0.4cm (7) $R_{U_iV_i}W_i$\par\hskip1cm
=${}^{F_i}R_{U_iV_i}W_i-(<{\text {grad}}\ f_i,\ {\text {grad}}\ f_i>/f_i^{2})(<U_i, W_i>V_i - <V_i, W_i>U_i)$\par\hskip1cm=${}^{F_i}R_{U_iV_i}W_i + 
\left( ({f_i'}^{+}u(t-p)+{f_i'}^{-}u(p-t))/{f_i}^2\right)(<U_i, W_i>V_i - <V_i, W_i>U_i).$
\par
\vskip10pt
{\it Proof}\par We will establish (3) and (7)\par
\hskip0.4cm For $X=X^{1}{{\frac{\partial }{\partial t}}}$,\ $Y=Y^{1}{{\frac{\partial }{\partial t}}}$ \par
since  \  $\nabla_{X}{\ {\text {grad}}\ f_i}$=$\nabla_{X}(-){\frac{\partial f_i}{\partial t}}{\frac{\partial }{\partial t}}$=$
-\left({\frac{\partial f_i}{\partial t}}X^{1}\nabla_{{\frac{\partial }{\partial t}}}^{\ {\frac{\partial }{\partial t}}}+ X^{1}{\frac{\partial }{\partial t}}\left({\frac{\partial f_i}{\partial t}}\right){\frac{\partial }{\partial t}}\right)$
\par\hskip2.6cm
=$-X^{1}\left(f_i''(t)+\delta_{p}(t)\left({f_i'}^{+}-{f_i'}^{-}\right)\right){\frac{\partial }{\partial t}}$\par
\hskip0.4cm (3) $H^{f_i}(X,Y)=\biggr\langle \nabla_{X}{\ {\text {grad}}\ f_i},Y \biggr\rangle$\par\hskip2.6cm
=$\biggr\langle -X^{1}\left(f_i''(t)+\delta_{p}(t)\left({f_i'}^{+}-{f_i'}^{-}\right)\right){\frac{\partial }{\partial t}} ,Y^{1}{{\frac{\partial }{\partial t}}}\biggr\rangle$\par\hskip2.6cm=$X^{1}Y^{1}\left(f_i''(t)+\delta_{p}(t)\left({f_i'}^{+}-{f_i'}^{-}\right)\right)$\par
\hskip1cm $R_{U_iX}Y = (H^{f_i}(X, Y)/f_i)U_i=\biggl\{\biggl\{X^{1}Y^{1}\left(f_i''(t)+\delta_{p}(t)\left({f_i'}^{+}-{f_i'}^{-}\right)\right)\biggr\}/f_i
\biggr\}V$\par
\hskip0.4cm (7) $R_{U_iV_i}W_i={}^{F}R_{U_iV_i}W_i-(<{\text {grad}}\ f_i,\ {\text {grad}}\ f_i>/f_i^{2})(<U_i, W_i>V_i$ \par\hskip2.6cm $ -<V_i, W_i>U_i)$\par\hskip2.4cm=${}^{F}R_{U_iV_i}W_i+\left( ({f_i'}^{+}u(t-p)+{f_i'}^{-}u(p-t))/{f_i}^2\right)(<U_i, W_i>V_i$\par\hskip2.6cm$ - <V_i, W_i>U_i).$\qed
\vskip15pt
{\bf Corollary 3.3}\ \ Let $M=B\times_{f_1} F_1\times...\times_{f_n} F_n$ be a multiply warped products with Riemannian curvature tensor $R$. If $X$, $Y\in \frak{L}(B)$, $U_i$, $V_i\in\frak{L}(F_i)$ where $1\leq i\leq n$, $d_i={\text {dim}}\thinspace F_i$ $f_i\in C^0(S)$,  then   \par
\hskip0.4cm  (1) ${\text {Ric}}(X,Y)$=$-\sum_{i=1}^n({d_i}/f_i) H^{f_i}(X, Y)$\par\hskip2.5cm
=$-\sum_{i=1}^n({d_i}/f_i)X^{1}Y^{1}\left(f_i''(t)+\delta_{p}(t)\left({f_i'}^{+}(p)-{f_i'}^{-}(p)\right)\right)$\par
\hskip0.4cm  (2) ${\text {Ric}}(X,U_i)$=0 \par
\hskip0.4cm (3) ${\text {Ric}}(U_i,V_i)$=${}^{F_i}{\text {Ric}}(U_i, V_i)$\par\hskip2.8cm$+<U_i, V_i>\biggl\{\left(f_i''(t)+\delta_{p}(t)({f_i'}^{+}-{f_i'}^{-})\right)/f_i$\par\hskip2.8cm$+(d_i-1)\left({f_i'}^{+}-{f_i'}^{-}\right)/f_i^{2}+\sum_{j\not=i} d_j{\frac{<{f_i'}^{+}-{f_i'}^{-},\ {f_j'}^{+}-{f_j'}^{-}>}{f_if_j}}\biggr\}$\par
\hskip0.4cm (4) ${\text {Ric}}(U_i, U_j)=0$\  for $i\not=j$,\par
\vskip10pt
{\it Proof}\par
We will establish (1) and (3)\par
\hskip0.4cm (1) ${\text {Ric}}(X,Y)$=$\sum_{M}\varepsilon_m<R_{XE_m}Y, E_m>$\par
\hskip2.55cm =$\sum_{F_i}\varepsilon_m<R_{XE_m}Y, E_m>$\par
\hskip2.55cm =$\sum_{F_i}\varepsilon_m<{\frac{-H^{f_i}(X,Y)}{f_i}}E_m, E_m>$\par
\hskip2.55cm =$-\sum_{F_i}{\frac{H^{f_i}(X,Y)}{f_i}}\varepsilon_m<E_m, E_m>$\par
\hskip2.55cm =$-\sum_{F_i}{\frac{H^{f_i}(X,Y)}{f_i}}\varepsilon_m^2$\par
\hskip2.55cm =$-\sum_{i=1}^n{\frac{d_i}{f_i}}H^{f_i}(X,Y)$\par
\hskip2.55cm =$-\sum_{i=1}^n({d_i}/f_i)X^{1}Y^{1}\left(f_i''(t)+\delta_{p}(t)\left({f_i'}^{+}(p)-{f_i'}^{-}(p)\right)\right)$\par
\hskip0.4cm (3) Since $\triangle\ f_i$ = ${\text {div(grad}}\ f_i)
= -  \biggl\{f_i''(t)+\delta_{p}(t)\left({f_i'}^{+}(p)-{f_i'}^{-}(p)\right)\biggr\},$\par

\hskip1cm${f_i^{\sharp}}={\frac{\triangle{f_i}}{f_i}}+({d_i}-1){\frac{<{\text {grad}}\ f_i,\ {\text {grad}}\ f_i>}{f_i^2}}$\par\hskip1.5cm
$= -\biggl\{f_i''(t)+\delta_{p}(t)\left({f_i'}^{+}(p)-{f_i'}^{-}(p)\right)\biggr\}/f_i
$\par$\hskip1.8cm
-(d_i-1)\biggl\{{f_i'}^{+}u(t-p)/f_i+{f_i'}^{-}u(p-t)/f_i\biggr\}^{2} $\par
therefore \par
\ \ \ \ \  
${\text {Ric}}(U_i,V_i)$=${}^{F_i}{\text {Ric}}(U_i, V_i)$\par\hskip2.5cm$+<U_i, V_i>\biggl\{\left(f_i''(t)+\delta_{p}(t)({f_i'}^{+}-{f_i'}^{-})\right)/f_i$\par\hskip2.5cm$+(d_i-1)\left({f_i'}^{+}-{f_i'}^{-}\right)/f_i^{2}+\sum_{j\not=i} d_j{\frac{<{f_i'}^{+}-{f_i'}^{-},\ {f_j'}^{+}-{f_j'}^{-}>}{f_if_j}}\biggr\}$\qed
\vskip15pt
Now consider the multiply warped products, $M=R^{1}\times_{f_1} F_1\times...\times_{f_n} F_n$ with the warping function $f_i$ on the spacelike hypersurface 
$\Sigma=\{(x_1,...,x_{n+1})|x_1\text{\ is constant}\}$. If $M$ has a metric $g=-dt^2+\sum_{i=1}^nf_i^2g_i$,\  metric of $F_i$ ( ${\text {dim}}\thinspace F_i=d_i$ ), is $C^{\infty}$ and symmetric.
\vskip15pt
{\bf Proposition 3.4}\ \ Let $M=B\times_{f_1} F_1\times...\times_{f_n} F_n$ be the multiply warped product space of  dimension $d=1+\sum_{i=1}^nd_i$ with the warping functions $f_i$ on the spacelike hypersurface $\Sigma=\{p\}\times F_1\times...\times F_n$. The metric components of $g=g^{L}\cup g^{R}$ are $C^{1}$ functions of the coordinate variables ie, $f_i\in C^1$ if and only if $[L_{\eta}]=(L_{\eta}^R-L_{\eta}^L)=0$ at each point on the spacelike hypersurface $\Sigma$. Here $\eta={\frac{\partial}{\partial t}}$
\vskip10pt
{\it Proof}
\par
We can get the second fundamental form as follow. 
For $M=B\times_{f_1} F_1\times...\times_{f_n} F_n$, consider the basis of $T_pM$.
 $$\Bigl\{{\frac{\partial}{\partial x_{1}}}, {\frac{\partial}{\partial {\mu}_{1\ 1}}},...,{\frac{\partial}{\partial {\mu}_{1\ d_1}}},...,{\frac{\partial}{\partial {\nu}_{n\ 1}}},...,{\frac{\partial}{\partial {\nu}_{n\ d_n}}}\Bigr\}=\Bigl\{{\frac{\partial}{\partial x_{1}}}, {\frac{\partial}{\partial x_{2}}},...,{\frac{\partial}{\partial x_{d}}}\Bigr\}$$
For $p\in\Sigma$, $X_p=a_2{\frac{\partial}{\partial x_{2}}}+a_3{\frac{\partial}{\partial x_{3}}}+,...,+ a_d{\frac{\partial}{\partial x_{d}}}\in T_{\Sigma}$, choose normal vector ${\eta_p}={\frac{\partial}{\partial x_{1}}}$ such that $g(X_p,\eta_p)=0$ for non-degenerate $g$\ \ where $p=(x_1, x_2,...,x_{d+1})$ and the $a_i$ are smooth. 

Since $$L_{\eta}X =-\nabla_X{\eta}=-\sum_{i=2}^{d+1}{\frac{a_i}{f_i}}{\frac{\partial f_i}{\partial x_{1}}}{\frac{\partial}{\partial x_i}}$$

therefore, we have\par
$$\Bigl[L_{\eta}\Bigr]=-\sum_{i=2}^{d+1}{\frac{a_i}{f_i}}\Bigl[{\frac{\partial f_i}{\partial x_{1}}}\Bigr]{\frac{\partial}{\partial x_i}}$$
$$\Bigl[L_{\eta}\Bigr]=0\Leftrightarrow \Bigl[{\frac{\partial f_i}{\partial x_1}}\Bigr]=0 \Leftrightarrow f_i \in C^1(\Sigma)\qed$$
\vskip15pt
\noindent{\bf IV. SCHWARZSCHILD SPACE-TIME AS A MULTIPLY WARPED PRODUCT}
\vskip10pt
In this section we will briefly discuss the interior Schwarzschild solution. We show how the interior solution can be written as a multiply warped product. \vskip10pt

A {\it Schwarzschild black hole} for the region $r<2m$ have the metric. 
$$ds^2=-\Bigl({\frac{2m}{r}}-1\Bigr)^{-1}dr^2+\Bigl({\frac{2m}{r}}-1\Bigr)dt^2+r^2d\Omega^2\eqno(4.1)
$$
Replacing $r$ with $\nu$ and $t$ with $\mu$, we have $0<\nu<2m$ and 
$$ds^2=-\Bigl({\frac{2m}{\nu}}-1\Bigr)^{-1}d{\nu}^2+\Bigl({\frac{2m}{\nu}}-1\Bigr)d\mu^2+{\nu}^2d\Omega^2\eqno(4.2)
$$
 Put $$d\mu^2=\Bigl({\frac{2m}{\nu}}-1\Bigr)^{-1}d{\nu}^2\eqno(4.3)$$
Integrating 
 $$d\mu=\sqrt{{\frac{\nu}{2m-\nu}}}d{\nu}$$
we obtain $$\mu=2m\thinspace {\text{cos}}^{-1}\Bigl(\sqrt{\frac{2m-\nu}{2m}}\Bigr)-{\sqrt{\nu(2m-\nu)}}+C=F(\nu)+C.\eqno(4.4)$$
Setting $C=0$ we obtain $F(\nu)=2m\thinspace {\text{cos}}^{-1}\Bigl(\sqrt{\frac{2m-\nu}{2m}}\Bigr)-\sqrt{\nu(2m-\nu)}$\par
This yields $$\lim_{\nu\to 2m}F(\nu)=m\pi, \ \ \ \lim_{\nu\to 0}F(\nu)=0  $$
Notice ${\frac{d\nu}{d\mu}}>0$ implies $F^{-1}$ is a well-defined function.\par
By using (4.4) we rewrite (4.2) as
$$ds^2=-d\mu^2+\Bigl({\frac{2m}{F^{-1}(\mu)}}-1\Bigr)d\nu^2+{F^{-1}(\mu)}^2d\Omega^2\eqno(4.5)
$$
Therefore we can write Schwarzschild space-time as a multiply warped products
$$ds^2=-d\mu^2+f^2_1(\mu)d\nu^2+f^2_2(\mu)d\Omega^2\eqno(4.6)$$
where $$f_1(\mu)=\sqrt{\Bigl({\frac{2m}{F^{-1}(\mu)}}-1\Bigr)}$$ and 
$$f_2(\mu)={F^{-1}(\mu)}$$\par
Clearly one may investigate the curvature of the interior metric(4.6) of Schwarzschild space-time as a multiply warped product. Furthermore, we have the following {\it{Ricci curvature}} on the multiply warped products.\par\vskip10pt
{\bf Corollary 4.1}\ \ Let $M$ be a multiply warped product $M=R^1\times_{f_1} R^1\times_{f_2} S^2$ with metric $ds^2=-d\mu^2+f^2_1(\mu)d\nu^2+f^2_2(\mu)d\Omega^2$(where $d\Omega^2=d\theta^2+{\text{sin}^2{\theta}}\thinspace d\phi^2$) for warping functions $f_1$, $f_2$.\par Then we have {\it{Ricci curvature}}\par
\hskip0.4cm  (1) ${\text {Ric}}({\frac{\partial}{\partial \mu}},{\frac{\partial}{\partial \mu}})$=${\text {R}}_{\mu \mu}$=${\text {R}}_{1 1}$=$-{\frac{f_1''}{f_1}}-{\frac{2f_2''}{f_2}}$\par
\hskip0.4cm  (2) ${\text {Ric}}({\frac{\partial}{\partial \nu}},{\frac{\partial}{\partial \nu}})$=${\text {R}}_{\nu \nu}$=${\text {R}}_{2 2}$=$f_1f_1''+{\frac{2f_1f_1'f_2'}{f_2}}$\par
\hskip0.4cm (3) ${\text {Ric}}({\frac{\partial}{\partial \theta}},{\frac{\partial}{\partial \theta}})$=${\text {R}}_{\theta \theta}$=${\text {R}}_{3 3}$=${\frac{f_1'f_2f_2'}{f_1}}+{f_2'}^2+f_2f_2''+1$
\par
\hskip0.4cm (4) ${\text {Ric}}({\frac{\partial}{\partial \phi}},{\frac{\partial}{\partial \phi}})$=${\text {R}}_{\phi \phi}$=${\text {R}}_{4 4}$=$\Bigl({\frac{f_1'f_2f_2'}{f_1}}+{f_2'}^2+f_2f_2''+1\Bigr){\text{sin\ }}{\theta}$
\par
\hskip0.4cm  (5) ${\text {R}}_{\text{mn}}=0$\  for ${\text m}\not={\text n}$\par
\vskip10pt

{\bf Remark}\ \ Let $M=(0,\ 2m)\times_{f_1}R^1\times_{f_2} S^2$ be given the metric $ds^2=-d\mu^2+f^2_1(\mu)d\nu^2+f^2_2(\mu)d\Omega^2$. Then $M$ is the {\it{Ricci flat}} with interior Schwarzschild metric if $\lim_{\mu\to 0}({f_2'(\mu)}^2+1)f_2(\mu)=2m$ and $\lim_{\mu\to 0}f_1(\mu)=\lim_{\mu\to 0}f_2'(\mu)$ \par
\vskip10pt
{\it Proof}\par
From corollary 4.1 we have ${\frac{f_1''}{f_1}}=-{\frac{2f_2''}{f_2}}$ and ${\frac{f_1''}{f_1}}=-{\frac{2f_2'}{f_2}}$ from ${\text {R}}_{1 1}=0$, ${\text {R}}_{2 2}=0$ respectively. \par
Substitute ${\frac{f_1'}{f_1}}={\frac{{\frac{f_1''}{f_1}}}{{\frac{f_1''}{f_1'}}}}={\frac{f_2''}{f_2'}}$ to ${\frac{1}{f_2}}\times {\text {R}}_{3 3}=0$, then we have ${\frac{f_1'f_2'}{f_1}}+{\frac{{f'_2}^2+1}{f_2}}+f_2''=0$ \par so $2f_2''+{\frac{{f_2'}^2+1}{f_2}}=0$ thus ${\frac{2{f_2}'f_2''}{{f_2'}^2+1}}+{\frac{f_2'}{f_2}}=0$. After integrating from $0$ to $\mu$ \par we have ${\text {ln}}\ ({f'_2(\mu)}^2+1)f_2(\mu)=\lim_{\mu\to 0}{\text {ln}}\ ({f'_2(\mu)}^2+1)f_2(\mu)$. \par Put $\lim_{\mu\to 0}({f'_2(\mu)}^2+1)f_2(\mu)=C$,   
${f'_2(\mu)}^2+1={\frac{C}{f_2(\mu)}}$ so ${f'_2}(\mu)=\sqrt{\Bigl({\frac{C}{f_2(\mu)}}-1\Bigr)}$.\par
From ${\frac{f_1'}{f_1}}={\frac{f_2''}{f_2'}}$  we have $f_1(\mu)={\frac{f_(\mu)}{f'_2(\mu)}}f'_2(\mu)=f'_2(\mu)$ with initial condition\par $\lim_{\mu\to 0}{\frac{f_(\mu)}{f'_2(\mu)}}=1$ 
that is, ${f_1}(\mu)=\sqrt{\Bigl({\frac{C}{f_2(\mu)}}-1\Bigr)}$
therfore ${f_1}(\mu)=\sqrt{\Bigl({\frac{2m}{f_2(\mu)}}-1\Bigr)}$ if $C=2m$.\par
Since $f_2(\mu)={F^{-1}(\mu)}=\nu$ and $f_1(\mu)=\sqrt{\Bigl({\frac{2m}{F^{-1}(\mu)}}-1\Bigr)}={\frac{d\nu}{d\mu}}$ 
we have (4.3) \qed
\vskip15pt
\noindent{\bf ACKNOWLEDGMENTS}
\vskip10pt
I wish to thank to John K. Beem for his invaluable guidence, encouragement and help. 
\vskip18pt
${}^1$ Bishop, R. L. and B. O'Neill
Manifolds of negative curvature
Trans. A.M.S.\par\hskip0.3cm 145,  1 (1969)
\vskip5pt
${}^2$ Beem, J. K., P. E. Ehrlich and K. Easley
{\it{Global Lorentzian Geometry, Second\par\hskip0.3cm edition}}
 Marcel Dekker Pure and Applied Mathematics, New York, 1996
\vskip5pt
${}^3$ Beem, J. K., P. E. Ehrlich and T. G. Powell 
 Warped product manifolds in\par\hskip0.3cm relativity, in selected studies : A Volume Dedicated to the Memory of Allbert\par\hskip0.3cm Einstein (T. M. Rassias and G. M. Rassias, eds) North-Holland, Amsterdam,\par\hskip0.3cm 41 (1982)
\vskip5pt
${}^4$ Beem, J. K. and P. E. Ehrlich
 Singularities, incompleteness, and the Lorentzian\par\hskip0.3cm distance function
 Math. Proc. Camb. Phil. Soc.
 85, 161 (1979)
\vskip5pt
${}^5$ O'Neill, B. {\it {Semi-Riemannian Geometry with Applications to Relativity}}\par\hskip0.3cm Academic Press Pure and Applied Mathematics, 1983      
\vskip5pt
${}^6$ Choquet-Bruhat. Y
Espaces-temps einsteiniens g\'{e}n\'{e}raux, chocs gravitationels\par\hskip0.3cm
 Ann. Inst. Henri Poincar\'{e} 8, 327 (1968)
\vskip5pt
${}^7$ Powell, T. G.
 Lorentzian manifolds with non-smooth metrics and warped\par\hskip0.3cm products
 Ph.D. Dissertation, University of Missouri-Columbia 1982
\vskip5pt
${}^8$ Taub, A. H. 
 Singular hypersufaces in general relativity
 J. Math. 1, 
 370 (1965)
\vskip5pt
${}^{9}$ Taub, A. H. 
Space-times with distribution valued curvature tensor
 J. Math.\par\hskip0.3cm Phys. 21, 1423 (1980)
\vskip5pt
${}^{10}$ Hoskins, R. F.  
{\it{Generalized Functions}}
Ellis Horwood limited, Oxford,\par\hskip0.4cm 1979      
\vskip5pt
${}^{11}$ Kanwal, R. P.
{\it{ Generalized Functions Theory and Technique}}(2nd ed.)
Birkh\"{a}user, \par\hskip0.4cm1997      
\vskip5pt
${}^{12}$ Lichnerowicz, A.
 Theor\'{e}or\'{i}es relativistes de la gravitation et de l'\'{e}lectromag\'{e}ticame
 \par\hskip0.4cm Masson et Cie, Paris, 1955
\vskip5pt
${}^{13}$ Smoller, J. and B. Temple 
 Shock waves near the Schwarzchild radius and\par\hskip0.4cm stability limits for stars
Physical Review D 55(12), 7518 (1997)
\vskip5pt
${}^{14}$ Flores, J, L. and M. S\'{a}nchez
Geodesic connectedness of multiwarped\par\hskip0.4cm spacetimes  preprint 1999 
\vskip5pt
${}^{15}$ Steven G. Harris. 
 Topology of the future chronological boundary: Universality \par\hskip0.4cm for spacelike boundaries Preprint 1999
\vskip5pt
${}^{16}$ Hawking, S. W. and F. R. Ellis
{\it{The Large Scale Structure of Space-times}}\par\hskip0.4cm
Cambridge University Press, Cambridge, 1973
\vskip5pt
${}^{17}$ Misner, C. W.,  K. S. Thorne and J. A. Wheeler 
Gravitation  W. H. Freeman \par\hskip0.4cm and Company New York, 1970
\enddocument
\end